\documentclass[aps,a4paper,twocolumn]{revtex4}
\usepackage{epsfig}
\usepackage{graphicx}
\usepackage{amsmath,amssymb,color,xcolor}
\usepackage[english]{babel}
\usepackage[colorlinks=true, allcolors=blue]{hyperref}
\usepackage{comment}
\usepackage{gensymb}
\usepackage{enumitem}

\newcommand{\Fp}{F_\mathrm{p}}
\newcommand{\Fc}{F_\mathrm{c}}

\renewcommand{\d}{\delta}

\parskip=\medskipamount

\begin{document}
\title{The role of zealots in the spread of linguistic traits}

\author{V.~Dornelas$^1$, C.~Anteneodo$^{2, 3}$, R.~Nunes$^1$, E.~Heinsalu$^1$, M.~Patriarca$^{1,4}$}

\affiliation{
$^1$  National Institute of Chemical Physics and Biophysics, Tallinn, Estonia; \\
$^2$ Department of Physics, Pontifical Catholic University of Rio de Janeiro, 
Rio de Janeiro, Brazil; \\
$^3$ National Institute of Science and Technology for Complex Systems, 
Rio de Janeiro, Brazil; \\
$^4$ Department of Cybernetics, Tallinn University of Technology, 19086 Tallinn, Estonia.
}

\date{\today}

\begin{abstract}
We investigate the diffusion of linguistic innovations on a fully connected network in order to understand the emergence of linguistic diversity.
We employ an agent-based dynamics based on the Axelrod model, where interactions between agents are driven by homophily and social influence, with the difference that we assume that all agents share a number of common features that ensure a finite probability of pairwise interaction.
We start from a homogeneous population and introduce zealots that act like agents spreading linguistic innovations, without being influenced by other agents.
We analyze how different factors, such as the degree of cohesion and number of zealots in different linguistic states, determine the linguistic configurations that populations can adopt and contribute to the possible emergence of a multi-linguistic community. 
The results are compared with those derived within the mean-field approximation.
\end{abstract}


\maketitle

\section{Introduction}
\label{sec:intro}

The concept of totally or partially inflexible agents has become a relevant element in the modeling of social systems. 
Stubborn agents~\citep{ramos2015does}, mass media or super-agents~\citep{Shibanai_2001}, and committed minorities~\citep{xie2011social} have been shown to be useful paradigms in describing a wide range of social phenomena involving the appearance of extreme views or behaviors.
Particularly, zealots are defined as entirely inflexible agents capable of modifying or influencing other agents without being affected by them. 
In the framework of language dynamics, zealots represent individuals or groups that resist linguistic changes due to a strong preference toward a specific language and its features (also called linguistic purism). 
The effects of zealots have been widely explored also in opinion dynamics \citep{mobilia2007role, meyer2024time, mobilia2015nonlinear, galam2007role} as well as in the binary naming game \citep{xie2011social, xie2012evolution, verma2014impact}, where their impact -- promoting consensus, polarization, or diversity -- depends on network topology and the rules of interaction.
Furthermore, the results of these works suggest that to drive the system to consensus, a certain minimum number of zealots is required. 

Zealotry has also been studied in the context of Axelrod’s dynamics by introducing super-agents \citep{Shibanai_2001} in the system. 
The model of Axelrod on the dissemination of culture is based on two key assumptions \cite{axelrod1997}: 1) homophily, i.e., the idea that individuals are more likely to interact with those with whom they are similar, and 2) social influence, i.e., the tendency that individuals become more similar as a result of interaction. 
Importantly, the first assumption leads to the situation where two agents who are extremely different, i.e., have no features in common, do not communicate. 
In the models that include super-agents, regular agents follow Axelrod's interaction rules, but can, in any case, interact with super-agents.

Notice that, though in his paper Axelrod studied how cultural traits spread through a population, he also pointed out that the very same model could be used to study languages. 
Thus far, however, the Axelrod model has been implemented only in very few studies of language dynamics \cite{Zankoc-2024,Hadzibeganovic-2009}. 
The goal of the present paper is to contribute to filling this gap.

In order to model language competition and evolution, 
we describe languages in terms of a given number of linguistic traits that can be transmitted between individuals through communication.
In the original Axelrod model, the assumption of homophily would imply that two very different languages should not influence each other.
In real situations, however, language contact processes are usually in action, accompanied by borrowing and other --- e.g. syntactic or phonetic --- types of mutual influence, even for very different languages. 

To model this situation, we introduce an extension to the Axelrod model by assuming that culturally heterogeneous agents can nevertheless communicate due to a set of common linguistic traits. 
The remaining traits are variable and can be modified through interactions between agents following the standard Axelrod dynamics.
The model thus obtained describes a type of linguistic diversity evolving around a common core and guarantees at any time a finite probability of communication between any pair of agents.
In practice, a landscape of ``dialects'', an idea already introduced by Axelrod\cite{axelrod1997}, can be realized by simply assigning to all agents a number of 
common traits that will behave effectively as immutable features, since they will not be changed by the Axelrod dynamics.
The level of common linguistic background is quantified by the degree of cohesion.  
Furthermore, we assume the presence of zealots that are in some --- but not all --- of the possible states allowed by the model.

In this study, the generalized Axelrod model is analyzed on a fully connected network, a topology known to promote homogenization in opinion dynamics.
Indeed, as shown in Ref.~\citep{dornelas2018impact}, increasing the connectivity and randomness of a network, the likelihood of reaching a consensus state becomes higher.
The model is studied in the mean field limit, comparing the results with the individual-based model simulations.
While in the studies of the Axelrod model random initial conditions are used, i.e., the initial traits of each agent are assigned randomly, in this paper, we will systematically explore also the effect of initial conditions corresponding to different fractions of agents in the various possible linguistic states. 

A central question addressed through the Axelrod modeling framework is how cultural or linguistic diversity can persist on a global scale despite the homogenizing effects of interactions.
Importantly, we show that the system converges either to consensus or to a heterogeneous coexistence state, where individuals are present not only in the states supported by the zealots but also in \textit{all} the other possible states obtained as combinations of the linguistic features of the zealots.
In other words, there is a minimal number of different zealot types, smaller than the number of different states, which ensures the maximal linguistic diversity of the population.

\section{Model description}
\label{sec:model}

We consider a system consisting of $N$ agents that interact pairwise through a dynamics similar to the one described by the Axelrod model.
Each agent represents either an individual or a small, linguistically homogeneous group of individuals, characterized by a specific linguistic state. 
We also assume that these agents can be either \textit{zealots}, i.e., agents who do not change, or \textit{regular agents}, who adapt their linguistic traits through interactions.
We analyze the system on a fully connected network, where every agent can interact directly with all other agents, mimicking a well-mixed population.

\subsection{Axelrod model}
\label{sec:Axelrod}

The Axelrod model~\citep{axelrod1997} characterizes each agent $i$ of the system ($i=1, \dots, N$) through a  state $\boldsymbol{\phi}_i$, defined as a vector of $F$ linguistic features,
$\boldsymbol{\phi}_i = (\phi_i^1, \phi_i^2, \ldots, \phi_i^F)$; notice that $F$ is constant in time.
Each feature  $\phi_i^\ell$ ($\ell = 1, \dots, F$) can assume one of $Q$ possible discrete values, capturing the linguistic variability. 
The probability of interaction between two agents $i$ and $j$ is determined by their degree of linguistic similarity, defined as
\begin{equation}
S_{ij} = 
        \frac{1}{F} \sum_{\ell=1}^F \delta(\phi_i^{\ell}, \phi_j^{\ell}) \, ,
    \label{eq:similarity_A}
\end{equation}
where $\delta(u,v)$ is the Kronecker delta. 
The degree of similarity can assume values between zero and one. 
When the agents share no common features ($S_{ij} = 0$),  the probability of their interaction is zero. 
When they are linguistically identical ($S_{ij} = 1$),  the probability of interaction is maximal, although such interactions do not change the agents' states.

\subsection{Generalized Axelrod model}
\label{sec:generalizedAxelrod}

In the present paper, we consider a variant of the Axelrod model to reflect a situation where communication between agents is always possible ($S_{ij} \neq 0$), based on the assumption that all agents -- regular agents and zealots -- share at least some common features. 
Thus, the state of each agent $i$ is defined by a vector $\boldsymbol{\phi}_i$ that is divided into two subsets of common and personal linguistic traits,
\begin{equation} \label{eq:phii}
\boldsymbol{\phi}_i =  \Big(
\underbrace{ \phi_{i}^1, \phi_{i}^2, \ldots, \phi_{i}^{\Fc} }_{\text{$\Fc$ common features}}, \underbrace{ \phi_{i}^{\Fc +1}, \ldots, \phi_{i}^{F}}_{\text{$\Fp$ personal features}} \Big) \, .
\end{equation}
The first subset consists of $F_{\rm c} > 0$ \textit{common} traits, which we consider constant in time. 
In the case of \textit{personal} traits $\Fp =F - \Fc$, the situation is different for regular agents and zealots. 
Namely, for \textit{regular agents}, the second subset consists of $\Fp$ \textit{variable} linguistic traits, each of which can take one of $Q$ possible variants. 
These variable traits are subject to social influence and can be copied between agents during the pairwise interactions, allowing them to evolve over time.
Instead, in the case of \textit{zealots}, the second subset consists of $\Fp$ \textit{fixed} linguistic traits that do not vary. 
However, because there can be different types of zealots in the system, the fixed linguistic traits can be different for different zealots.

In the case of languages, some linguistic features are more important than others and therefore, in calculating the degree of similarity within the analysis of a linguistic database, these features are given a larger statistical weight --- for applications to real linguistic datasets see~\citep{Polian2014VariacionDiasistema,Leonard2015ModelingAggregates,Leonard-2017,Patriarca2020LanguagesTime}. 
Furthermore, traits can often be understood as complex features composed of strongly interdependent sub-traits that tend to change together.
Also, in this case, a hierarchical structure can be formalized by assigning a specific weight to each trait, thus quantifying their relative importance or complexity.

To reflect these considerations, we define a weight vector $\boldsymbol{p} = (p_1, p_2, \ldots, p_F)$, where each $p_\ell$ corresponds to the weight attributed to feature $\ell$. 
In contrast to the model of Ref.~\citep{kalinowska2023weighted}, where the weights for traits are different for each agent (agents may prioritize different cultural features), we assume that the linguistic weights are fixed. 
Correspondingly, the weighted degree of linguistic similarity between two agents $i$ and $j$ is defined as
\begin{equation}
S_{ij} = 
        \frac{\sum_{\ell} p_{\ell} \, \delta(\phi_i^{\ell}, \phi_j^{\ell})}{\sum_{\ell} p_{\ell}} \, .
    \label{eq:similarity_h}
\end{equation}
In the following, we assume that all common traits have the same weight $p_{\rm c} > 0$ and all personal traits have the same weight $p_{\rm p} >0$. 
Thus, for our model, Eq.~\eqref{eq:similarity_h} becomes,
\begin{equation}
    \label{eq:similarity}
    S_{ij} = \frac{p_{\rm c} F_{\rm c} + p_{\rm p}\sum_{\ell} \delta( \phi_{i}^{\ell}, \phi_{j}^{\ell})}{p_{\rm c} F_{\rm c}+ p_{\rm p}\Fp } \, .
\end{equation}

In order to quantify the influence of common traits on language dynamics, we introduce the \textit{degree of cohesion} $f$, which measures the linguistic homogeneity in the population and is defined as the weighted ratio between the number of common features $F_c$ and the number of personal features $\Fp$,
\begin{equation}
    f \equiv \frac{p_{\rm c} F_{\rm c}}{p_{\rm p} \Fp} \, .
    \label{eq:f-def}
\end{equation}
We also introduce the Hamming distance $d$ between two agents, defined as the number of differing personal traits,
\begin{equation}
    d = \Fp - \sum_{\ell=1}^{\Fp} \delta( \phi_{i}^{\ell}, \phi_{j}^{\ell} )\, ,
    \label{eq:Hdistance}
\end{equation}
which can vary between $d = 0$ (when agents are identical) and $d = \Fp$ (when all the personal features are  different). 
Using the degree of cohesion (\ref{eq:f-def}) and the Hamming distance (\ref{eq:Hdistance}), one can express the degree of similarity (\ref{eq:similarity}) between two agents $i$ and $j$ differing by a Hamming distance $d$, as follows,
\begin{equation}
    S_d = \frac{f + (1 - d/\Fp)}{f + 1} \, .
    \label{eq:s_d}
\end{equation}

Having many common features with lower weight is equivalent to having a few common features with higher weight, meaning that the relevant parameter is the degree of cohesion, $f$. 
Since the model assumes that $F_{\rm c} > 0$ and $p_c > 0$, the cohesion factor is always strictly positive, $f > 0$, which implies that $S_d > 0$, i.e., there is always a finite probability of communication.
Notice that the cohesion $f$ is a continuous parameter $f \in (0,\infty)$ measuring the linguistic overlap between different states.

\begin{figure}[t!]
    \centering
    \includegraphics[width=0.85\linewidth]{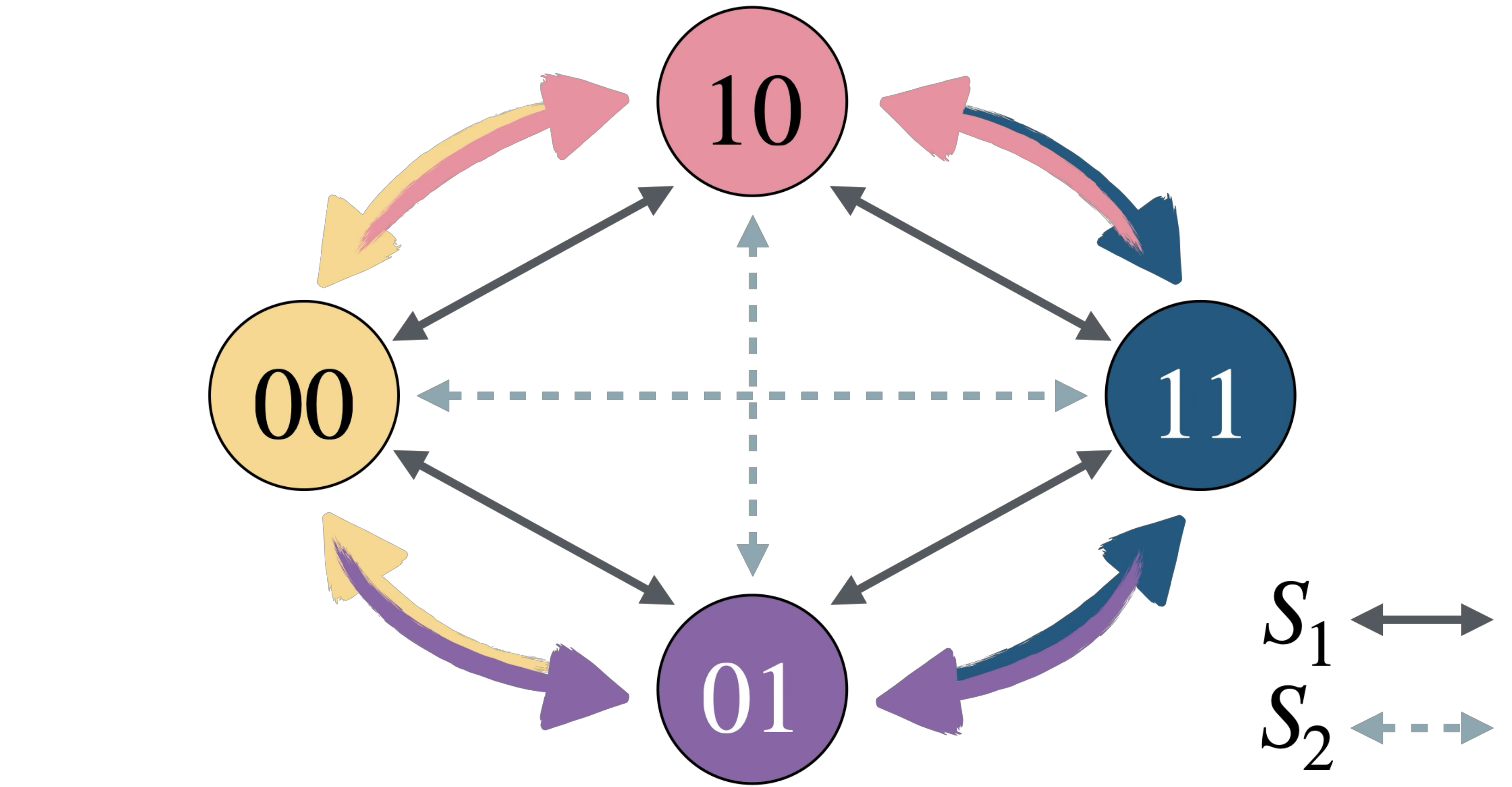}
    \caption{
        Diagram of the four linguistic states of the model with $\Fp = 2$ and $Q=2$ (only the personal traits are shown).
        The thin arrows represent interactions between agents with linguistic distance $d=1$ and degree of similarity $S_1$ (continuous lines) and $d=2$, $S_2$ (dotted lines), with $S_d$ defined in Eq.~(\ref{eq:s_d}).
        The interaction between agents in the same state ($d = 0$, $S_0$, not shown) does not change their states.
        The ticker external arrows represent the possible transitions between different states.
    }
    \label{fig:sketch}
\end{figure}

In the following, we consider a system in which agents have $\Fp = 2$ distinct personal traits with variability $Q = 2$, so that there are a total of $Q^{\Fp} = 4$ distinct linguistic states. 
The possible values of each trait are denoted by $0$ and $1$.
To simplify the notation, we label the four possible linguistic states as $\phi \in \{00, 01, 10, 11\}$, referring only to the personal traits without indicating explicitly their degree of overlap, see Fig.~\ref{fig:sketch}. 
We will refer to the states 00 and 11 as \textit{pure states} and to the states 01 and 10 as \textit{hybrid states}.
As illustrated in Fig.~\ref{fig:sketch}, interactions are possible between two agents in any state, but an agent can make a transition only to another state at a distance $d=1$. 
The probabilities of interaction between agents are determined by the possible values of the degree of similarity, i.e. by $S_d$ for $d = 0, 1, 2$.

\begin{figure}[t!]
    \centering
    \includegraphics[width=0.95\linewidth]{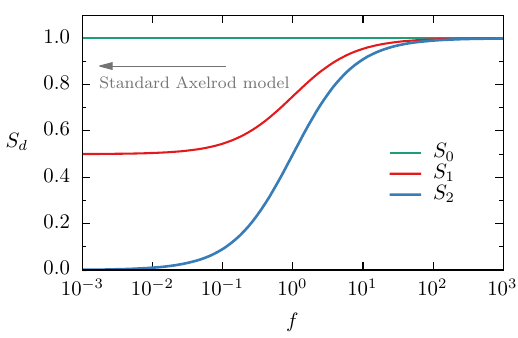}
    \caption{ Degree of similarity, $S_d$, 
    between two individuals differing by $d$ traits as a function of the degree of cohesion, $f$; $\Fp = 2$. 
    }
    \label{fig:fc/fv}
\end{figure}

The behavior of $S_d$ as a function of the degree of cohesion, $f$, is illustrated in Fig.~\ref{fig:fc/fv} for the different values of the distance $d$. 
As can be seen from Eq.~(\ref{eq:s_d}) and Fig.~\ref{fig:fc/fv}, when the interacting individuals are linguistically identical ($d = 0$), the degree of similarity reaches its maximum value, $S_0 \equiv S_{\rm max} = 1$.
In the opposite case, when all the personal features differ ($d = \Fp$), the degree of similarity is minimal, $S_{\Fp} \equiv S_{\rm min} = f/(1+f)$.
In the limit $f \gg 1$, the influence of personal traits becomes negligible and $S_d \to 1$, regardless of $d$. 
Instead, for $f \ll 1$, the influence of common traits becomes negligible and the original Axelrod model is recovered in the limit $f \to 0$.

The structure of the model for $\Fp=2, Q=2$ schematized in Fig.~\ref{fig:sketch} resembles other models of language and cultural competition~\citep{castellano2000nonequilibrium,PATRIARCA2012, Patriarca2020LanguagesTime}, such as the Minett and Wang model \citep{Minett2008a}, the basic naming game model \citep{Baronchelli_2006}, or the agent-based model \citep{Castello_2006}, where the transition from a ``pure state'' to another ``pure state''  cannot take place directly, but only through some intermediate state. 
Here, the role of intermediate states is played by the union of the two ``hybrid'' states 01 and 10.
Notice that in language competition models, the intermediate state corresponds to the bilingual state, and in this case, one can distinguish between two types of bilinguals on the basis of the mother tongue \citep{Scialla-2023a}, so that the three-state model is actually a four-state model as in the present case.

\subsection{Microscopic agent-based approach}
\label{sec:ABM}
The agent-based algorithm, which regulates the microscopic  dynamics, proceeds as follows:

\begin{enumerate}[leftmargin=*]
    
    \item At the initial time $t=0$, each agent is assigned a linguistic state.
   
    \item During each time step, an interaction process takes place, consisting of $N$ pairwise interactions. 
    For each interaction:
    \begin{itemize}[leftmargin=*]
        \item Two agents, $i$ (the ``hearer'') and $j$ (the ``speaker''), are randomly selected from the population.
        \item If agent $i$ is a zealot, 
        nothing happens and the next pair of agents is extracted.
        \item Instead, if agent $i$ is a regular agent, the probability of interaction determined by $S_d$ is computed from Eq.~\eqref{eq:s_d}.\\
        If $d=1$, then the different feature $\phi_j^\ell$ of agent $j$ is copied into the corresponding position of the hearer's vector, replacing $\phi_i^\ell(t)$.\\
        If $d=2$, then one of the speaker's personal features, $\phi_j^\ell$, is randomly selected and copied in the hearer's vector.\\
        Such a copying process increases the similarity between agents $i$ and $j$.
            
    \end{itemize}

    \item After all $N$ interactions at time $t$ are completed, the time is incremented as $t \to t + 1$, and the next interaction process begins.

\end{enumerate}
This algorithm is implemented in the numerical simulations presented in Sec.~\ref{sec:zealots}.

\subsection{Mean-field approximation}
\label{sec:MF}

To complement the microscopic approach, we derive the mean-field equations that describe the macroscopic dynamics of the system. 
The mean-field approach 
provides an alternative framework for describing the dynamics in terms of the population sizes associated with each linguistic state and allows to carry out the systematic analytical and numerical investigation of how zealots influence the dynamics and the outcome of the system's evolution.
However, while this formulation offers intuitive insights at the population level,  it becomes increasingly impractical for larger values of $\Fp$ and/or $Q$, since the number of possible linguistic states grows exponentially as $Q^{\Fp}$.

Notice that previous mean-field studies of the Axelrod dynamics have often described the system in terms of bonds, i.e., the fraction of neighboring agent pairs with a given overlap or similarity~\citep{castellano2000nonequilibrium, vazquez2007non, de2009effects}.
Instead, in the present model, where all agents can always communicate with each other, we reformulate the corresponding mean-field equations in terms of the population fractions in different states.

The fraction of regular agents in each state $\phi \in \{00, 01, 10, 11\}$ at time $t$ is $n_\phi(t)$.
The fraction of zealots in state 00 is $m_{00}$, and $m_{11}$ in state 11; it is assumed that there are no zealots in the hybrid states, i.e., $m_{01}=m_{10}=0$.
We introduce the variables $m$ and $\d m$ representing, respectively, the sum and the difference of the zealot fractions in pure states,
\begin{equation}
\begin{aligned}
\label{eq:mm}
    m = m_{00} + m_{11} \, ,\\
    \d m = m_{11}-m_{00} \, .
\end{aligned}
\end{equation}
The total fraction of the agents in the pure state $\phi$, including both regular agents and zealots, is denoted as $n_{\phi}^{\star}(t) \equiv n_{\phi}(t) + m_{\phi}$.
The population fractions satisfy at any time $t$ the normalization condition
\begin{align}
\label{eq:cons}
&n_{00}(t) + n_{01}(t) + n_{10}(t) + n_{11}(t) + m \nonumber\\
&\equiv n_{00}^{\star}(t) + n_{01}(t) + n_{10}(t) + n_{11}^{\star}(t) = 1 \, .
\end{align}
The fractions of zealots, $m_{00}$ and $m_{11}$, are constant in time. 
Instead, the fractions of regular agents in state $\phi$, $n_{\phi}(t)$, change in time depending on the probabilities of interaction between different linguistic states. 
The probability that a regular agent $i$ in state $\phi$ interacts with a randomly chosen agent $j$ in state $\phi'$, that differs from $\phi$ by a Hamming distance $d=1$, is given by $S_1 n_{\phi'}^{\star}$. 
As a result of the interaction, agent $i$ goes from state $\phi$ to state $\phi'$.
The probability that a regular agent $i$ in state $\phi$ interacts with a randomly chosen agent $j$ in state $\phi''$ differing from state $\phi$ by $d = 2$, is $S_2 n_{\phi''}^{\star}$. 
Because during each interaction only one feature can change, the agent $i$ cannot go from state $\phi$ directly to state $\phi''$. 
Instead, as a result of changing one of the features, agent $i$ can go to one of the two possible intermediate states -- denoted as $\phi_1'$ and $\phi_2'$ -- both of them occurring with probability $S_2 n_{\phi''}^{\star}/2$. 
The scheme of the possible interactions and transitions is shown in Fig.~\ref{fig:sketch}.

Considering all the interactions that cause a transition from state $\phi$ to state $\phi'$, the corresponding transition rate is,
\begin{equation} 
    T_{\phi \to \phi'} = S_1 n_{\phi'}^{\star} + \frac{S_2}{2} n_{\phi''}^{\star} \, ,
\end{equation}
while $T_{\phi \to \phi''} = 0$.
The mean-field equation describing the time evolution of agent fractions in each state is derived by summing all possible incoming and outgoing transitions. 
This yields for the dynamical equation the compact form
\begin{equation}
    \dot{n}_\phi = \sum\limits_{\phi'} \big( T_{\phi' \to \phi}\, n_{\phi'} - T_{\phi \to \phi'}\, n_\phi \big)  \, .
\end{equation}
Writing explicitly, the resulting system of equations reads
\begin{equation}
    \begin{aligned}
    \dot{n}_{00}(t) = & S_1  m_{00}(n_{01}+n_{10})- S_2 m_{11} n_{00}     
            \\& - S_2(n_{00}n_{11} - n_{01}n_{10}) \, ,  \\
    \dot{n}_{01}(t) = & - S_1 (m_{00}+m_{11})n_{01} + \frac{S_2}{2} (m_{11} n_{00} + m_{00} n_{11}) 
            \\& + S_2(n_{00}n_{11} - n_{01}n_{10}) \, , \\
    \dot{n}_{10}(t) = & - S_1(m_{00}+m_{11})n_{10} + \frac{S_2}{2} (m_{11} n_{00} + m_{00} n_{11}) 
            \\& + S_2(n_{00}n_{11} - n_{01}n_{10}) \, , \\
    \dot{n}_{11}(t) = & S_1 m_{11}(n_{01}+n_{10})  - S_2 m_{00} n_{11} 
            \\& - S_2(n_{00}n_{11} - n_{01}n_{10}) \, .
    \label{eq:meanfield}
    \end{aligned}
\end{equation}

Notice that the transitions of regular agents between states at a distance $d = 1$ take place at the same rate in the two opposite directions and thus cancel out in the mean field limit.
Transitions induced by the interactions between regular agents and zealots in states at $d=1$, instead, result in the terms proportional to $S_1$ in the dynamical equations (\ref{eq:meanfield}).
Transitions due to interactions between agents in states at a distance $d=2$ would be impossible in the basic Axelrod model, since in the absence of common features, the probability of interaction would be zero. 
In the present model, instead, they are possible.
In fact, in the case of interactions between two regular agents, these interactions are the only ones that contribute to the dynamics. 
Thus, the mean-field dynamics is dominated by interactions between agents in states at the largest linguistic distances $d=2$, a fact that unexpectedly contradicts the homophily hypothesis of the Axelrod model.

The mean-field equations \eqref{eq:meanfield} are invariant under the exchange of the variables $n_{01}$ and $n_{10}$. 
In the following, we assume that $n_{01}(0) = n_{10}(0)$, which implies that $n_{01}(t)\equiv n_{10}(t)$ at any time $t$.
This observation allows us to reduce the dimensionality of the system by introducing the new variable
\begin{equation}
    z(t) \equiv n_{01}(t) + n_{10}(t) \, ,
    \label{eq:z(t)}
\end{equation}
which represents the total fraction of individuals in the hybrid states 01 and 10. 
Notice that also in the model of Ref.~\cite{Scialla-2023a}, in the microscopic dynamics, one distinguishes between two types of bilinguals defined by the first language, but not at the macroscopic level.

Furthermore, we introduce the variable
\begin{equation}
    w(t) \equiv n_{11}^\star(t) - n_{00}^\star(t)\,,
    \label{eq:w(t)}
\end{equation}
which quantifies the asymmetry between the populations in the pure states 00 and 11 at time $t$, incorporating the contribution of zealots.  

Rewriting Eqs.~\eqref{eq:meanfield} in terms of the new variables $z(t)$, $w(t)$ and incorporating the constraint (\ref{eq:cons}), we obtain the following system of two coupled equations,
\begin{subequations} \label{eq:zw}
\begin{align}
    \dot{w}(t) &= \frac{S_2}{2} \, \left[ \d m\,(1-z) - m w \right] + S_1 \d m\, z  \,, \label{eq:zw-a} \\
    \dot{z}(t) &= \frac{S_2}{2} \, \left( 1 - m + \d m\,w - w^2 - 2z+mz\right)   
             -  S_1 m z \, .
\label{eq:zw-b}
\end{align}
\end{subequations}

Using the normalization condition and the definitions of the variables $w$ and $z$, Eqs.~\eqref{eq:z(t)}-\eqref{eq:w(t)}, we can express the pure state-populations $n_{00}$ and $n_{11}$ and the hybrid-state total population $n_{01} + n_{10}$ in terms of the $w$ and $z$ variables,
\begin{equation}
\label{eq:inv}
\begin{aligned}
    &n_{00} = (1 - z - w)/2 \, ,\\
    &n_{01} = z/2 \, , \\
    &n_{10} = z/2 \, ,\\
    &n_{11} = (1 - z + w)/2 \, .
\end{aligned}
\end{equation}
In the $w\,z$--plane, the domain accessible to the system can be found from the constraints $n_{00}\geq 0, n_{11}\geq 0$, and $n_{01}+n_{10}\geq 0$, which, using the equations above, read $z \le 1 - w, z \le 1 + w, z \geq 0$.

In the following, since Eq.~\eqref{eq:zw} does not admit a simple analytical solution, we numerically solve it using the Runge-Kutta method to examine how the presence of zealots and their different fractions affects the system's dynamics in the mean-field limit. We also compare the results with the ones obtained from numerical simulations of the agent-based model.

\begin{figure*}[t!]
    \centering
    \includegraphics[width=\textwidth]{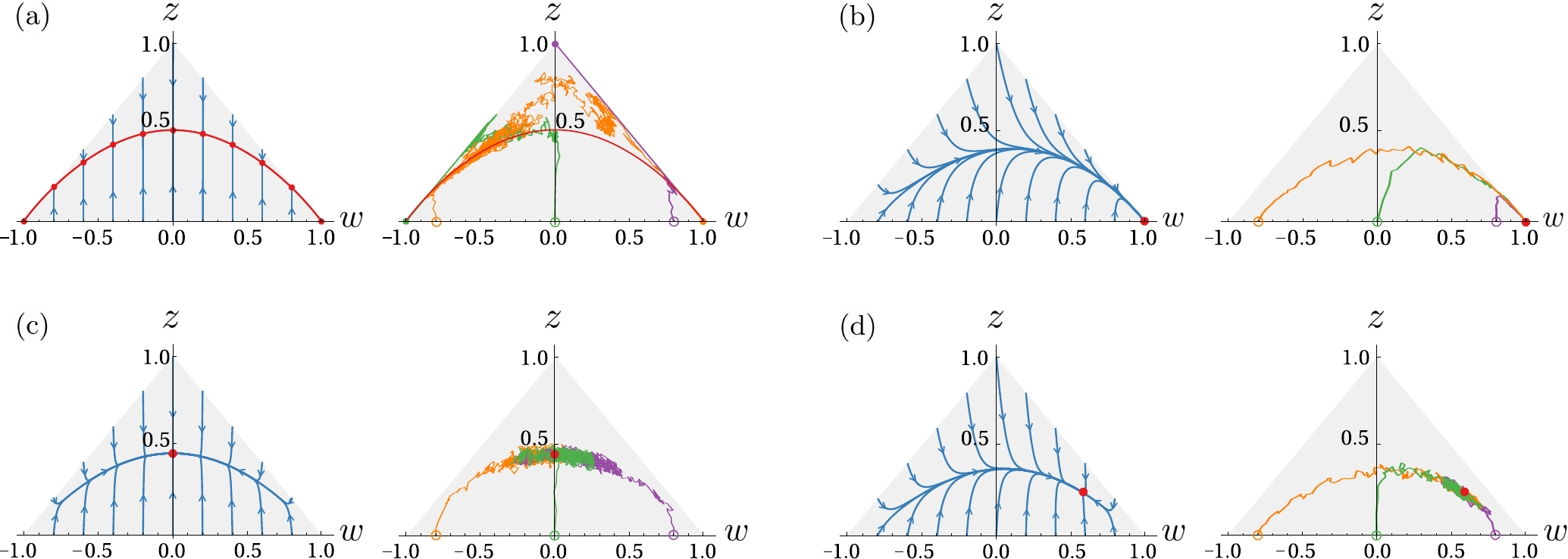}
    \caption{ Phase portrait of the generalized Axelrod model in the $wz$-plane. Each panel shows, on the left, the mean-field dynamics and, on the right, single realizations from agent-based simulations with $N = 10^3$ agents, for three different initial conditions marked by open dots. 
    The triangular shaded area represents the phase space accessible to the system, defined by the constraints $z\ge0, z \le 1 - w, z \le 1 + w$:
    (a) no zealots, $m=0$;
    (b) zealot of one type, $m_{00}=0$ and $m_{11}=0.08$;
    (c) two zealots, symmetrical case, $m_{00}=m_{11}=0.02$;
    (d) two zealots, asymmetrical case, $m_{00}=0.04$ and $m_{11}=0.08$.
    Red dots mark the fixed points of the system under mean-field dynamics. 
    The red curve in panel (a) is formed by fixed points. 
    The degree of cohesion is $f=0.5$. }
    \label{fig:PP_wz}
\end{figure*}

\section{Results}
\label{sec:zealots}

In the standard Axelrod dynamics, the system typically evolves toward one of two final configurations: either full consensus or fragmented cultural regions that no longer interact, depending on the network structure and the system parameters~\citep{Tucci_2016}. 
As our model ensures that the linguistic similarity $S_{ij}$ between any two individuals $i$ and $j$ is always strictly positive, a fully static fragmented state cannot occur. 
The only true absorbing configuration is consensus, where $S_{ij} = 1$ for any $i$ and $j$.
In the presence of zealots, however, the system can reach a dynamic stationary regime, characterized by the sustained coexistence of agents in different linguistic states.

In the following, we analyse the system's dynamics by comparing the predictions from the mean-field approximation with the results from agent-based simulations on fully connected networks. 
While the mean-field equations describe the average macroscopic behavior of the system, the agent-based model captures stochastic fluctuations arising from finite-size effects, providing insight into how noise can influence the system’s dynamics and the final outcome.

Because the model assumes that only two types of zealots are possible, we investigate the system in the absence of zealots, in the presence of one type of zealots, and in the presence of two types of zealots. 
In the latter case, we address the symmetrical situation when the fractions of the zealots in states 00 and 11 are equal, $m_{00} = m_{11}$, and the asymmetrical situation when the fractions are different, $m_{00} \neq m_{11}$.
The phase portraits in the $wz$-plane for these four representative cases of zealot fraction configurations are depicted in Fig.~\ref{fig:PP_wz}. 
In the same figure, we draw for comparison also the single realizations from agent-based simulations with different initial conditions.

\subsection{Absence of zealots}

In the absence of zealots ($m = 0$), the mean-field equation, Eq.~\eqref{eq:zw-a}, indicates that $\dot{w}(t) = 0$, implying that the variable $w$ remains constant over time; thus, its equilibrium value is $w_s = w(0)$. 
In this case, one can see from the phase portrait in Fig.~\ref{fig:PP_wz}(a) that the mean-field trajectory evolves vertically and reaches the isocline defined by $\dot{z} = 0$, given by a curve that consists entirely of fixed points, $z_s = (1 - w_s^2)/2$, see Eq.~(\ref{eq:zw-b}).
This curve crosses the accessible domain going from the point $(w_s, z_s) = (-1, 0)$ corresponding to 00-consensus to the other 11-consensus point $(w_s,z_s) = (1, 0)$ crossing the point $(w_s, z_s) = (0, 0.5)$ where the population is equally shared among the four states 00, 01, 10, 11.
Thus, in this regime, the final state of the system depends on the initial conditions, due to the absence of external biases.
The system evolves spontaneously, driven solely by local interactions among individuals. 

The comparison in Fig.~\ref{fig:PP_wz}(a) reveals that agent-based simulations exhibit a very different dynamics from those associated to the mean-field trajectories.
Specifically, while the mean-field trajectories approach the line of fixed points (red curve), these points do not correspond to stationary states of any agent-based model realizations.
Instead, agent-based simulations invariably reach one of the possible consensus states, located at the corners of the triangular domain, where all agents have adopted identical linguistic traits.
Which particular consensus state $\phi$ is reached is a random event characterized by a certain probability given by the mean field equilibrium population fraction $n_\phi$.

For example, let us consider the following initial conditions: $w(0) = 0$ and $z(0) = 0$, which correspond to the population consisting of regular agents only,  initially distributed symmetrically between states 00 and 11, i.e., $n_{00} = n_{11} = 0.5$ and $n_{01} = n_{10} = 0$. 
The mean-field trajectory evolves then to the fixed point $w_s = 0, z_s = 0.5$, corresponding to equal fractions of agents, $n_\phi = 0.25$, in any state $\phi = 00,01,10,11$, i.e., there is a symmetrical coexistence at equilibrium of all four linguistic states.
Correspondingly, for each agent-based realization, the system will reach full consensus in one of the four states, with equal probabilities.

Instead, if the system starts from asymmetric initial conditions, for example, from $w(0) = 0.7$ and $z(0) = 0$, the mean-field approach predicts convergence towards a state favoring the dominant initial state, namely the fixed point with coordinates $w_s = 0.7$ and $z_s = 0.255$. 
From Eqs.~\eqref{eq:inv}, these values correspond to equilibrium population fractions in the pure states given by $n_{00} = 0.0225, n_{11} = 0.7225$ and a total population in the hybrid states $n_{01} + n_{10} = 0.255$. 
As in the previous example, these values also reflect the probabilities that the agent-based dynamics reaches consensus in each of the corresponding states.

\begin{figure}[t!]
    \centering
    \includegraphics[width=0.9\columnwidth]{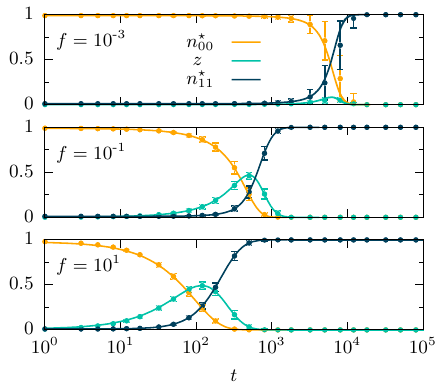}
    \caption{
    One type of zealots: $m_{11} = 10^{-2}$, $m_{00} = 0$. 
    Temporal evolution of $n_{00}^\star(t)$, $z(t)$, and $n_{11}^\star(t)$ for three different values of $f$. 
    Solid lines represent the mean-field solutions; dots with error bars correspond to results from the agent-based model averaged over $10^3$ realizations and with $N = 10^4$ agents.
    }
    \label{fig:ev-1z}
\end{figure}
%

\subsection{One type of zealots}

In the presence of only one type of zealots, the system is subjected to a single bias that causes all trajectories to converge to a unique stable global consensus state, both in the mean-field and in the microscopic limit.
If all zealots are for example in state 11, i.e., $m_{11} > 0$ and $m_{00} = 0$, then $m = \d m > 0$ and the only stable fixed point of Eqs.~\eqref{eq:zw} is $w_s = 1, z_s = 0$, corresponding to consensus in the state 11; $n_{00}^\star\equiv n_{00} = 0$, $n_{11}^\star=1$, and $n_{01} = n_{10} = 0$. 
This situation is illustrated in the phase portrait depicted in Fig.~\ref{fig:PP_wz}(b).
If instead all zealots are in state 00, $m_{00} > 0$ and $m_{11} = 0$, this leads to the stable fixed point $w_s=-1, z_s=0$, corresponding to consensus in the state 00, $n_{00}^\star=1, n_{11}$, and $n_{01}=n_{10}=0$. 
Importantly, the consensus emerges independently of the initial conditions, which aligns with earlier works in opinion dynamics, such as the voter model on one- and two-dimensional lattices \citep{mobilia2003does}, where it has been found that even a small committed minority can drive the entire population toward consensus. 
This contrasts with phenomena observed in other models \citep{xie2011social,xie2012evolution,verma2014impact,mobilia2015nonlinear}, where a minimum threshold of committed individuals is necessary to trigger consensus.

In Fig.~\ref{fig:ev-1z}, we depict some examples of time evolution of the population fractions for different degrees of linguistic cohesion $f$, comparing the results from the mean-field approximation and the agent-based model. 
The comparison shows an excellent correspondence; as expected from mean-field theory, the agreement is the better the larger is the population size $N$, but it also improves for larger values of $f$.
Furthermore, the standard deviation represented by error bars decays with the population size $N$ following $1/\sqrt{N}$.
In the simulations of Fig.~\ref{fig:ev-1z}, it is assumed that initially there are only zealots in state 11 and that all regular agents are in state 00.
The curves show that for higher values of $f$, the transient is characterized by a higher fraction $z(t)$ of intermediate hybrid states and the convergence toward the linguistic state supported by the zealots is significantly accelerated compared to the lower values of $f$.
In contrast, for lower values of $f$, a smaller fraction of agents in the intermediate states appears during the time evolution, and the system undergoes a more direct transition from the 00 to the consensus 11 state, visible as a steeper step. 

\subsection{Two types of zealots}

In the mean-field approximation, when both types of zealots are present, the system does not converge to a consensus either in 00 or 11. 
Instead, the competition between the zealots leads to a unique global fixed point that does not depend on initial conditions and corresponds to the stable coexistence of all four linguistic states; see panels (c) and (d) in Fig.~\ref{fig:PP_wz} for equal or different fractions of the two types of zealots, respectively. 
This means that in a system where there are four linguistic groups, the influence of only two committed linguistic groups is sufficient to sustain maximal diversity in the final linguistic landscape. 
Notice that it is not important that the zealots are in the pure states defined above; they could also be in the hybrid states. It only matters that they are at the maximal linguistic distance from each other ($d = 2$).
It can be expected that, also in more general Axelrod models with $\Fp$ features and trait variability $Q$, the presence of different agents that are maximally different from each other can induce an exploding recombinant diversification process, encompassing all the $Q^{\Fp}$ possible states.
This follows from the feature-wise influence characterizing the model, i.e., the possibility that agents can pass a single feature to each other, analogously to a horizontal gene transfer process \citep{Castellano-2009a}.
If there are no zealots, such high diversity may eventually be reabsorbed, depending on the system parameters.
However, if zealots are present and are maximally different from each other, this diversity will persist indefinitely, as in the case discussed here. 

Importantly, in the case of the agent-based model, in the presence of two types of zealots, the system does not converge to a static configuration; instead, as can be seen from Fig.~\ref{fig:PP_wz}(c) and (d), it fluctuates around the equilibrium points, corresponding to a dynamic steady state.

Considering the symmetric case, when the two zealot communities have equal sizes, i.e., $m_{00} = m_{11} \equiv m/2$ and $\delta m = 0$, and using $2S_1/S2 -1= 1/f$ [compare Eq.~\eqref{eq:s_d}], the mean-field equations \eqref{eq:zw} predict a fixed point corresponding to a dynamic coexistence between the four cultural states, with stationary fractions

\begin{subequations} 
\label{eq:2z-smf}
\begin{align} 
    w_s &= 0 \, , \label{eq:2z-smf-ws} \\
    z_s &= \frac{1 - m}{2 + m\left(1 + 1/f\right)} < \frac{1}{2} \, .  \label{eq:2z-smf-zs}
\end{align}
\end{subequations}
These values correspond to equal populations in the pure states 
\begin{equation}
    n_{00} = n_{11} = \frac{1 - z_s}{2} = \frac{1 + 2m(1 + 1/2f)}{4 + 2m(1 + 1/f)} \, .
    \label{eq:n_symm}
\end{equation}
For $f \to 0$, also $z_s \to 0$; instead, in the limit $f \gg 1$, $z_s \to (1 - m)/(2 + m)$.

\begin{figure}[t!]
    \centering
    \includegraphics[width=0.9\columnwidth]{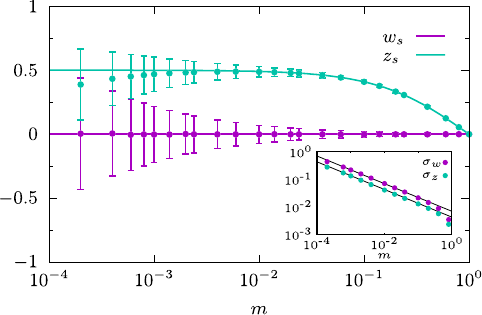}
    \caption{
    Two types of zealots: $m_{00} = m_{11}$. 
    Stationary values of $w_s$ and $z_s$ \textit{versus} $m$ for $f=1$. 
    Solid lines represent mean-field predictions; dots with error bars correspond to results from the agent-based model averaged over $10^3$ realizations and with $N = 10^4$ agents. 
    Inset: the standard deviations $\sigma_w$ and $\sigma_z$ \textit{versus} $m$. 
    }
    \label{fig:sym_wz}
\end{figure}

\begin{figure}[t!]
    \centering
    \includegraphics[width=0.9\columnwidth]{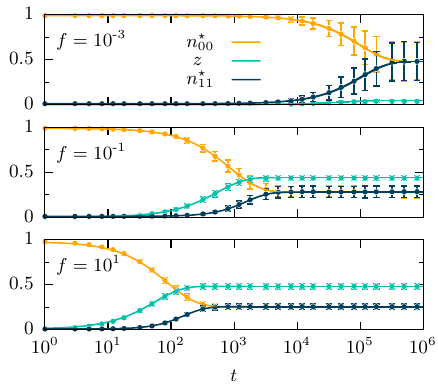}
    \caption{
    Two types of zealots: $m_{00} = m_{11} = 10^{-2}$. 
    Temporal evolution of $n_{00}^\star(t)$, $z(t)$, and $n_{11}^\star(t)$ for three different values of $f$.
    Solid lines represent the mean-field solutions; dots with error bars correspond to results from the agent-based model averaged over $10^3$ realizations and with $N = 10^4$ agents.
    }
    \label{fig:ev-2z}
\end{figure}

\begin{figure}[t!]
    \centering
    \includegraphics[width=0.9\columnwidth]{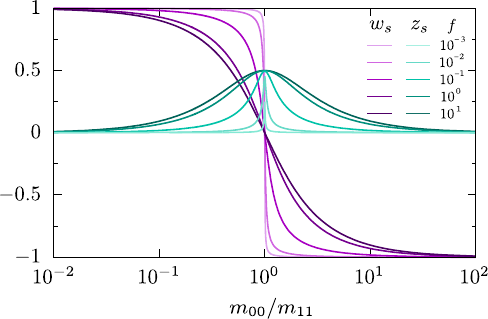}
    \caption{Two types of zealots: $m_{00} \neq m_{11}$. Stationary values of $w_s$ (lilac curves) and $z_s$ (green curves) \textit{versus} $m_{00}/m_{11}$ for various values of $f$ (mean-field approximation); $m_{00} = 10^{-3}$. 
    } 
    \label{fig:asym_wz}
\end{figure}
\begin{figure}[t!]
    \centering
    \includegraphics[width=0.9\columnwidth]{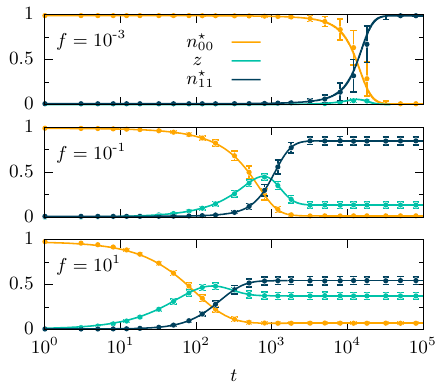}
    \caption{
    Two types of zealots: $m_{11} = 10^{-2}$ and $m_{00} = m_{11}/2$. 
    Temporal evolution of $n_{00}^\star(t)$, $z(t)$, and $n_{11}^\star(t)$ for three different values of $f$. 
    Solid lines represent the mean-field solutions; dots with error bars correspond to results from the agent-based model averaged over $10^3$ realizations and with $N = 10^4$ agents.
    }
    \label{fig:ev-2z-asym}
\end{figure}



In Fig.~\ref{fig:sym_wz}, the mean-field prediction (\ref{eq:2z-smf}) for $w_s$ and $z_s$ as functions of $m = m_{00} + m_{11}$ (solid lines) is compared with the results from the agent-based model (dots with error bars). 
The comparison shows a strong agreement between the two methods.
However, as indicated above, in the case of the agent-based model, the system does not converge to a static configuration but fluctuates around the equilibrium values $w_s$ and $z_s$. 
The error bars represent the standard deviation $\sigma_w$ and $\sigma_z$, capturing the variability of $w$ and $z$ over time and across different realizations.
These error bars, and thus $\sigma_w$ and $\sigma_z$, are larger for smaller values of $m$, and decrease as $m$ increases and, correspondingly, the influence of zealots becomes larger, suppressing the fluctuations and stabilizing the system around the mean-field predictions $w=w_s$ and $z=z_s$, given by Eqs.~\eqref{eq:2z-smf}. 
This stabilizing effect is quantitatively illustrated in the inset of Fig.~\ref{fig:sym_wz}, which shows that $\sigma_w$ and $\sigma_z$ decay approximately as $m^{-1/2}$ -- a scaling characteristic of finite-size systems subject to stochastic fluctuations. 
This dependence on $m$ was also observed in the dynamics of the voter model with zealots \citep{mobilia2007role, chinellato2015dynamical}.
The corresponding time evolution of the population fractions for different values of $f$ is depicted in Fig.~\ref{fig:ev-2z}, comparing the results from the mean-field approximation and the agent-based model.

\begin{figure*}[t!]
    \centering
    \includegraphics[width=0.99\textwidth]{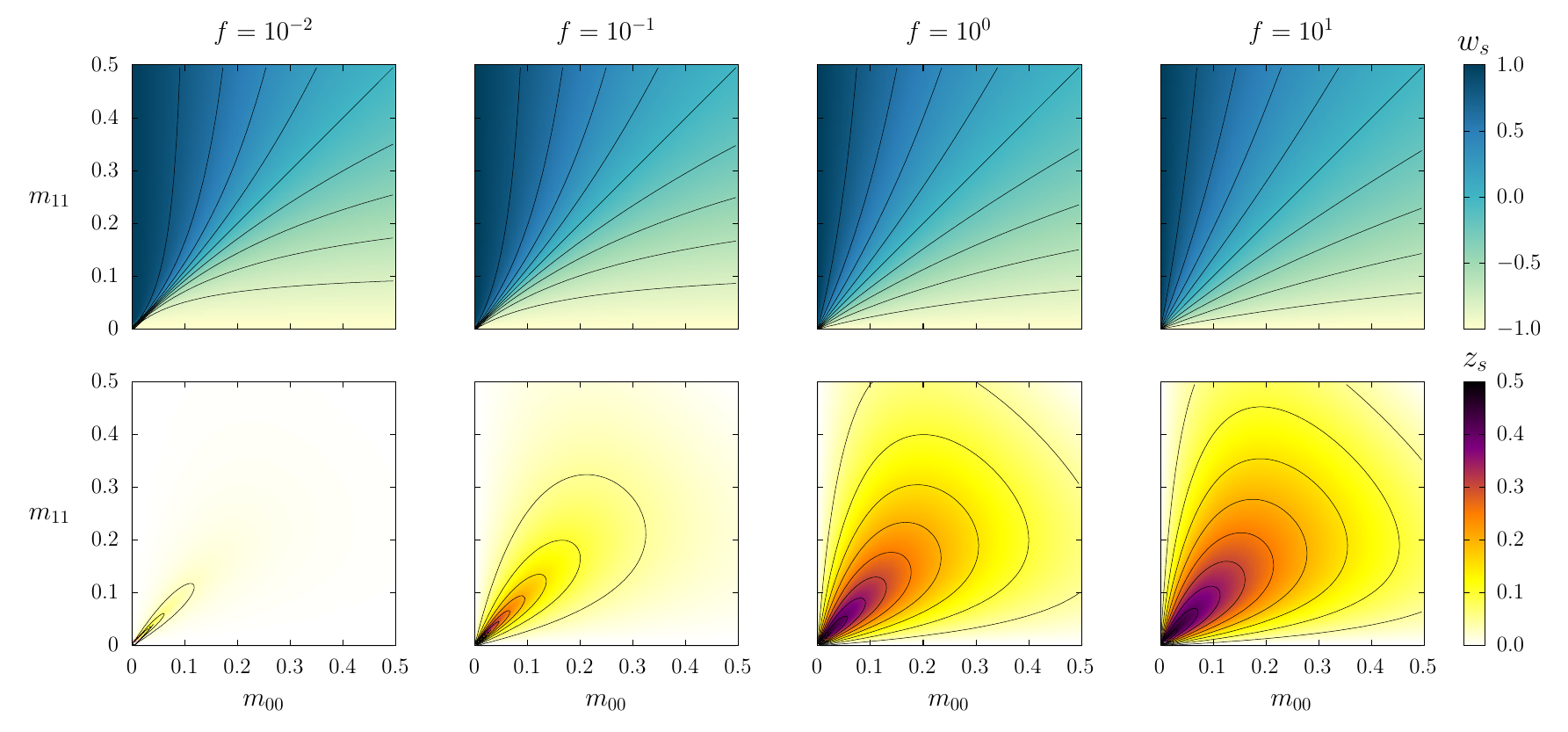}
    \caption{Two types of zealots. Phase diagrams of $w_s$ (upper panels) and $z_s$ (lower panels) in the $m_{00} \, m_{11}$-plane for different values of $f$ (mean-field approximation). }
    \label{fig:PD}
\end{figure*}

When the fractions of zealots are unequal, $m_{00}\neq m_{11}$, this leads to an equilibrium in which the linguistic state with the larger fraction of zealots becomes dominant.
Figure~\ref{fig:asym_wz} shows how equilibrium values of $w_s$ and $z_s$ depend on the ratio $m_{00}/m_{11}$ for a fixed $m_{00}$ and for various levels of cohesion $f$. 
As can be seen from this figure, for highly imbalanced cases -- either $m_{00} \gg m_{11}$ or $m_{11} \gg m_{00}$ -- the system approaches a near-consensus state in the dominant linguistic state, indicated by $|w_s|\approx 1$ and $z_s \approx 0$.
For smaller values of asymmetry, the coexistence of all four linguistic states is possible, corresponding to $z_s \neq 0$. 
A higher degree of cohesion enhances the system’s resilience to the dominance of a single linguistic state, allowing linguistically diverse states to persist over a broader range of zealot ratios.  
The fraction of agents in the intermediate hybrid states, $z_s$, reaches a maximum for the symmetric case $m_{00}/m_{11} = 1$; the value of the maximum is determined by Eq.~(\ref{eq:2z-smf-zs}).
The results in Fig.~\ref{fig:asym_wz} indicate also that beyond a certain level of cohesion, further increase of $f$ has minimal impact on the macroscopic behavior of the system [see also Eq.~(\ref{eq:2z-smf-zs}) for the symmetric fractions of the zealots].

Although in Fig.~\ref{fig:asym_wz} we have only presented stationary values obtained from numerical solution of the mean-field approximation, systematic agent-based simulations were also performed and confirm excellent agreement with these results. 
The good agreement between the mean-field approximation and the agent-based model results is demonstrated in Fig.~\ref{fig:ev-2z-asym}, where the time evolution of the system with $m_{00} \neq m_{11}$ is presented.
The comparison between Figs.~\ref{fig:ev-1z}, \ref{fig:ev-2z}, and \ref{fig:ev-2z-asym} demonstrates that the fluctuations in the agent-based model are the largest for the system with equal zealot fractions, $m_{00} = m_{11}$, and they decrease with the increase of the bias caused by the asymmetry between the zealot fractions.
This highlights how balanced zealot influence promotes higher dynamical variability, whereas greater asymmetry leads the system toward more stable and predictable consensus-like states.
Furthermore, for different fractions of zealots, the total fraction $z(t)$ of agents in hybrid states passes a maximum, as in the case with a single type of zealot (c.f. Fig.~\ref{fig:ev-1z}), stabilizing at a constant value afterwards.
If the fractions of the two types of zealots are equal, then $z(t)$ does not pass any maximum, but goes monotonously to an asymptotic value (see Fig.~\ref{fig:ev-2z}).

In order to have a better understanding and a exahustive overview of how the fractions of the two types of zealots and the degree of cohesion determine the equilibrium state of the system, we present in Fig.~\ref{fig:PD} the phase diagrams of $w_s$ and  $z_s$ in the $m_{00} \, m_{11}$-plane for different values of $f$. 

At low levels of cohesion, $f \ll 1$ ($f=10^{-2}$ in Fig.~\ref{fig:PD}), in most of the $m_{00} \, m_{11}$-plane one has that $z_s\approx 0$, meaning that the majority of the population is in the pure states 00 or 11, apart in a small region around the origin, $m_{00},m_{11}\ll 1$.
As the cultural cohesion $f$ increases, a larger and larger fraction of the population is in the hybrid states 01 and 10, however, never exceeding the maximum fraction $z= 0.5$ indicated in Eq.~\eqref{eq:2z-smf-zs}. 
At all values of the cohesion $f$ --- also for $f \gg 1$ ($f = 10$  in Fig.~\ref{fig:PD}) --- a fraction of the population remains in the pure state of the dominant zealot group, namely in the regions where the presence of the two types of zealots is highly asymmetrical, i.e., $m_{00} \ll m_{11}$ or $m_{00} \gg m_{11}$.
Furthermore, as discussed already above, beyond a certain value of the cohesion $f$, the changes in the phase diagrams become negligible, indicating a saturation effect.

\begin{figure}[h!]
    \centering
    \includegraphics[width=0.9\columnwidth]{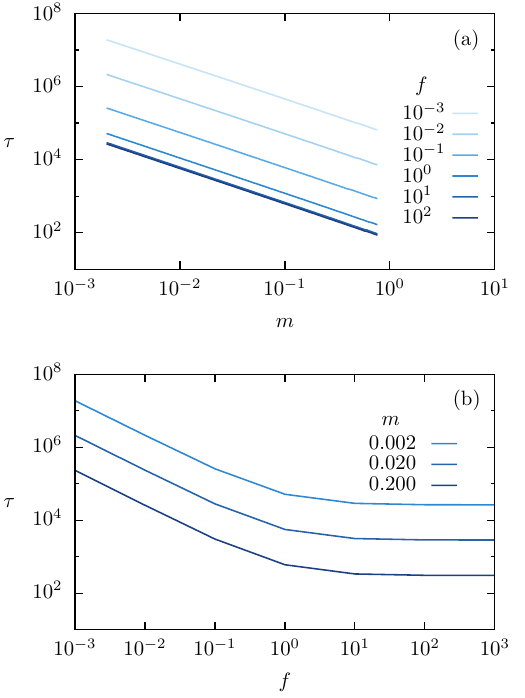}
    \caption{Two types of zealots: $m_{00} = m_{11}$. Convergence time $\tau$ \textit{versus} $m$ [panel (a)] and $f$ [panel (b)] for various values of $f$ and $m$, correspondingly. }
    \label{fig:Time}
\end{figure}

Finally, as any evolution process is inherently dynamical, it is worth studying the timescales over which the dynamics of the system stabilize. 
A useful measure of such a timescale is the \textit{convergence time} $\tau$, defined as the time required for the system to reach a stable, steady-state configuration.
Results in Fig.~\ref{fig:Time} show that both the total fraction $m$ of the two types of zealots and the degree of cohesion $f$ significantly influence the convergence time $\tau$. 
For a fixed value of $f$, the convergence time decreases with increasing zealot fraction $m$, following a power-law relationship $\tau \propto 1/m^\alpha$, with an exponent $\alpha \approx 0.96$ [see Fig.~\ref{fig:Time}(a)],
meaning that higher zealot fractions substantially accelerate the system's convergence. 
Instead, for a fixed value of $m$, the convergence time $\tau$ reaches a plateau around $f \approx 10$ [see Fig.~\ref{fig:Time}(b)], related to the saturation effect of the fixed traits, mentioned above, beyond which a further increase in the degree of cohesion $f$ does not significantly reduce $\tau$. 
The figure also demonstrates that for smaller values of $f$, the system exhibits slower dynamics. 
This slower evolution results from the dominance of personal linguistic traits before the system enters the saturation regime.

\section{Conclusion  and discussion}
\label{sec:conclusion}

In this paper, we have introduced and studied a language dynamics model. This model aimed to describe how the evolution of a linguistic landscape is determined by the level of common linguistic background, heterogeneity, and variability of the linguistic traits of the agents interacting in a fully-connected network, and by the presence of zealots. 
In a linguistic framework, zealots represent individuals who maintain fixed linguistic traits. 
The proposed model is based on the Axelrod model \cite{axelrod1997} with the difference that it assumes the presence of a number $\Fc$ of immutable traits common to all agents, which ensures that two agents can always interact; see also Ref.~\cite{Morales-2025}. 
The agents are either regular agents, whose remaining $\Fp$ personal traits can change as a result of interactions, or zealots. 
Moreover, each linguistic trait can be assigned a statistical weight, reflecting a different relevance or an internal hierarchical structure.

In the present paper, we have focused on a simple system with a common linguistic background, associated with a degree of cohesion $f>0$, and $F_p=2$ personal traits that can assume $Q = 2$ values, 0 or 1, thus resulting in four possible combinations 00, 01, 10, and 11. 
The zealots are assumed to be only in a state with 00 or 11.
The original motivation underlying this model and the particular choice of $F_p = 2$, $Q = 2$ comes from the study of the spreading of some phonetic features among the Friulian languages.
While Friulan languages share many common features with each other, they also present a set of distinct linguistic traits that are found in different combinations in different regions~\cite{Leonard-2009}. 
However, in the present paper, we have focused on the features of the mathematical model only, and its application to Friulan languages will be presented in a forthcoming paper.

To understand the role of zealots on system dynamics and final outcome, we have investigated three scenarios: the absence of zealots, the presence of a single type of zealots, and the presence of two distinct types of zealots.
The results of the mean-field approximation have been compared to the numerical simulations of the agent-based microscopical model.

We have observed that in the absence of zealots, there is a qualitative difference between the evolution predicted by the mean-field approximation and the many-agent model.
In the mean-field approximation, there is a continuum of fixed points characterized by the coexistence of the four linguistic states; which equilibrium configuration is reached depends on the initial conditions.
Instead, in the many-agent dynamics, 
the system evolves to the consensus in one of the four possible linguistic states, with a probability
given by the corresponding equilibrium population fraction obtained from the mean-field equations.

In the presence of a single type of zealots, the system reaches consensus in the state favored by the zealots, both in the agent-based simulations and in the mean-field limit, demonstrating that in a well-mixed population even a small proportion of zealots in a single state can impact the system and lead to consensus without the need of a critical mass.
This contrasts with models like the $q$-voter model~\citep{mobilia2015nonlinear} or the binary naming game model~\citep{xie2011social, xie2012evolution, verma2014impact}, which typically require the zealot population to exceed a certain threshold fraction in order to trigger a transition to consensus.

When there are two different types of zealots, the system's behavior is mainly shaped by the ratio between the zealot population sizes $m_{00}$ and $m_{11}$, similarly to the case of the symmetrical and asymmetrical voter model~\citep{mobilia2007role,chinellato2015dynamical}. 
Importantly, we have shown that the presence of only two committed linguistic groups at the maximal linguistic distance is sufficient to sustain maximal diversity in the system, i.e., one observes the coexistence of all sub-populations.
However, though in the asymmetric case, $m_{00} \ne m_{11}$, the agents still populate all four possible states, the system exhibits a bias toward the majority-supported state, reaching a near-consensus configuration in the limit of strong asymmetry.

The model discussed here in the framework of language dynamics lends itself to various straightforward generalizations that can be applied in other fields, such as opinion dynamics and cultural dynamics.
The particular values for $F_p$ and $Q$ are also relevant in social systems \cite{Flache-2011,GENZOR2015200}.
The hierarchical traits \citep{Madsen2015} used in this work can be expanded and adapted to different problems characterized by cultural states with hierarchical traits, each trait having a different weight and correlations with other traits.
In a cultural spreading framework, the set of common traits can represent ``cultural universals'', i.e., cultural traits or practices found in all known human cultures, or cultural cores shared by different groups, which facilitate mutual communication \cite{Morales-2025}.
Possible applications range from the diversification of religious traditions, social norms, and myths to the development and interaction between diversified technologies and political systems. 

Models of these types would open new possibilities for a quantitative modeling-based exploration and interpretation of cultural databases characterized by a complex structure \cite{Valori-2012, Liu-2022}.
One should also incorporate realistic network structures or multiplexity in the modeling of cultural states, as in the model of Battiston et al. \cite{battiston2017layered}, as well as strategic adaptation. 
Finally, although the model investigated here assumes that committed individuals (zealots) remain fixed in their states indefinitely, it would be valuable to investigate scenarios where their influence diminishes over time.

\section*{Acknowledgments}

This work was supported by the Estonian Research Council through Grant PRG1059.	
CA acknowledges the partial financial support (311435/ 2020-3) of 
 Conselho Nacional de Desenvolvimento Cient\'{\i}fico e Tecnol\'ogico (CNPq), Brazil, 
and (CNE E-26/204.130/2024) Funda\c c\~ao de Amparo \`a Pesquisa do Estado do Rio de Janeiro (FAPERJ), Brazil, as well as (finance code 001) Coordena\c c\~ao de Aperfei\c coamento de Pessoal de N\'{\i}vel Superior (CAPES), Brazil.



\appendix
\renewcommand{\thesection}{\Alph{section}}
\renewcommand\thefigure{A\arabic{figure}}    

\setcounter{section}{0}
\setcounter{figure}{0}  


\vspace{1cm}

\end{document}